\documentclass[a4paper,11pt]{article}
\pdfoutput=1

\usepackage{jinstpub} % for details on the use of the package, please

\usepackage{lineno}
\usepackage{soul}
\title{CONCERTO: Readout and control electronics}

\author[a,1]{O. Bourrion,\note{Corresponding author}}
\author[a]{C. Hoarau,}
\author[a]{J. Bounmy,}
\author[a]{D. Tourres,}
\author[a]{C. Vescovi}
\author[a]{J.-L. Bouly,}
\author[a]{N. Ponchant,}
\author[c]{A. Beelen,}
\author[b]{M. Calvo,}
\author[a]{A. Catalano,}
\author[b]{J. Goupy,}
\author[c]{G. Lagache,}
\author[a]{J.-F. Mac{\'{\i}}as-P{\'{e}}rez,}
\author[a]{J. Marpaud,}
\author[b]{A. Monfardini.}

\affiliation[a] {Univ. Grenoble Alpes, CNRS, Grenoble INP\textsuperscript{$\dagger$}, LPSC-IN2P3, 38000 Grenoble, France \\
$\dagger$ Institute of Engineering Univ. Grenoble Alpes}
\affiliation[b] { CNRS, Univ. Grenoble Alpes, Grenoble INP\textsuperscript{$\dagger$}, Institut Néel, 38000 Grenoble, France}
\affiliation[c] { Aix Marseille Univ., CNRS, CNES, LAM, Marseille, France}
\emailAdd{olivier.bourrion@lpsc.in2p3.fr}

\abstract{
The CONCERTO spectral-imaging instrument was installed at the Atacama Pathfinder EXperiment (APEX) 12-meter telescope in April 2021.
It has been designed to look at radiation emitted by ionised carbon atoms, [CII], and use the "intensity Mapping" technique to set the first constraints on the power spectrum of dusty star-forming galaxies.
The instrument features two arrays of 2152 pixels constituted of Lumped Element  Kinetic Inductance Detectors (LEKID) operated at cryogenic temperatures, cold optics and a fast Fourier Transform Spectrometer (FTS).
To readout and operate the instrument, a newly designed electronic system hosted in five microTCA crates and composed of twelve readout boards and two control boards was designed and commissioned.
The architecture and the performances are presented in this paper.
}


\begin{document}
\maketitle
\flushbottom

%%%%%%%%%%%%%%%%%%%%%%%%%%%%%%%%%%%%%%%%%%%%%%%%%%%%%%%%%%%%%%%%%%%%%%%%%%%%%%%
\section{Introduction}
CONCERTO is a millimeter-wave low spectral resolution imaging-spectrometer with an instantaneous field-of-view of 18.6\, arcmin diameter operating in the range 130-310\,GHz \cite{Lagache2020,monfardini2021concerto}.
It is installed at the Atacama Pathfinder EXperiment (APEX) 12-meter telescope located at 5100\,m above sea level on the Chajnantor plateau.
The primary goal of CONCERTO is to observe the radiation emitted by ionised carbon atoms using the \emph{intensity Mapping} technique to set the first constraints on the power spectrum of dusty star-forming galaxies. Furthermore, it will open a new window towards large mapping speed low resolution spectroscopy for the study of galaxy clusters via the Sunyaev-Zel'dovich (SZ) effect \cite{sz}.

The instrument is composed of a dilution cryostat operating at 60\,mK, which features two arrays of 2152 pixels made of Lumped Element Kinetic Inductance  Detectors (LEKID) \cite{Day2003,Baselmans2012} and cold optics. 
To minimize the number of cryostat feedthroughs from the sensors to the warm electronics, each pixel array is constituted of only six transmission lines coupled to the LEKIDs. This amounts for about 360 detectors per feed-line.
In acknowledgment of the unavoidable fabrication dispersion, which can cause KID self resonant frequency overlapping, an average frequency separation of 2.5\,MHz was set by design between each LEKID of a given feedline.
Indeed, in the past few years we have put some efforts in investigating a post-processing (trimming) technique allowing to individually adjust the resonances position. 
Despite the good results \cite{shu_disp} we concluded that this approach is somewhat impractical for our Aluminium thin films, in particular for an instrument with planned lifetime of the order of several years.
Consequently, this leads to the requirement of having a 1\,GHz bandwidth readout electronics dedicated to each feedline.

The cryostat is coupled to a room-temperature Martin-Puplett Interferometer (MpI) \cite{MARTIN1970105} that is designed to obtain a spectral resolution up to 1.2\,GHz. 
To avoid atmospheric drifts during a single interferogram, this fast Fourier Transform Spectrometer (FTS) is designed to achieve four full interferograms per second and per pixel.
Thus, to properly sample the interferograms, the readout electronics shall sample the LEKID at a sampling rate of about 4\,kHz and the MpI movements must be controlled synchronously with the acquisition system.

As the last system constraints, i.e. the available room in the APEX, the reduced heat dissipation at high altitude due to lower convection and the requirement to have an easy remote operation and maintenance. 
Indeed the cabin drastically limit the available space for the electronic, the space available was $\rm 600 \times 200 \times 300\,mm^3$ for the entire CONCERTO readout, housekeeping and control of the C-cabin elements.
Therefore, most of the warm electronics are directly integrated in the instrument chassis.
This paper describes the readout and control electronics used for CONCERTO.

%%%%%%%%%%%%%%%%%%%%%%%%%%%%%%%%%%%%%%%%%%%%%%%%%%%%%%%%%%%%%%%%%%%%%%%%%%%%%%%
\section{Electronics system description}

The setup required to instrument a single resonator line is composed of a dedicated readout board operated at room temperature, which injects an excitation frequency comb in the cryostat and measures the modified returning signal.
The frequency comb has each of its tones tuned to the LEKID self-resonant frequencies and has to be generated in the 1.5-2.5\,GHz range.
In the cryostat, the excitation signal is fed down to the low temperature stages.
In order to attenuate the thermal noise coming from the warmer stages, we have installed at 4\,K an attenuator of -20\,dB. 
A further (distributed) attenuation of around -6\,dB is provided by the stainless steel (lossy) injection cables running from the 4\,K down to base temperature. 
This ensures the injected thermal noise is lower than the cold amplifier input noise.
NbTi superconducting coaxial cables transfer the LEKID output signal with virtually no losses to the 4\,K stage.
The signal is first amplified with a Low Noise Amplifier (LNA) installed in the same 4\,K stage\footnote{Arizona State University 0.5-3\,GHz; Gain: 30\,dB. Noise T=5\,K. Input 1\,dB Compression (minimum): -36\,dBm. Typical Power diss. 10\,mW.}, and then again with a second LNA amplifier at the 50\,K stage\footnote{MITEQ AFS3-02000400-08-CR-4 2-4\,GHz. Gain: 30\,dB. Noise T < 50\,K. Output 1\,dB Compression (minimum): +5\,dBm. Typical power diss. 50\,mW .}, for a total amount of roughly +60\,dB. 
Stainless steel cables link the 50\,K stage and the room temperature output.
The overall electrical gain of each radio-frequency line to and from the room temperature electronics has been measured and is confirmed to be about +10\,dB. 
Fixed attenuators at 300\,K are used to finely adjust the overall power to the available dynamic range of the ADC. 
The core principle of the readout system is similar to the one used in the NIKA2 experiment \cite{Bourrion_2016}.
Indeed, the excitation frequency comb is generated at baseband in the electronics using coordinate rotation digital computer (CORDIC) and the returning frequency comb is down-converted and analyzed by channelized Digital Down Converters (DDC) that provide In-phase (I) and Quadrature (Q) components. 
These components are used to compute each tone amplitude and phase.

Due to the limited space available at APEX, a large part of the warm electronics is directly integrated in the instrument chassis (see \cite{Lagache2020} for details on the instrumental setup).
CONCERTO needs (i) twelve KID readout boards to instrument the low and high frequency arrays , each composed of six resonator lines, (ii) one board dedicated to Martin-Puplett Interferometer management named  Motor Controller and Martin-Puplett Monitor (MCMPM) and (iii) one board devoted to the Cryostat Positioning System (CPS).
These boards were designed in the Advanced Mezzanine Card (AMC) format and are hosted in five compact micro Telecom Computer Architecture (microTCA) crates (Vadatech VT899).
Each crate has a dimension of $\rm 127\,mm \times 311\,mm \times 260\,mm$.

\begin{figure}[hbtp]
\centering
\includegraphics[angle=0,width=0.99\textwidth]{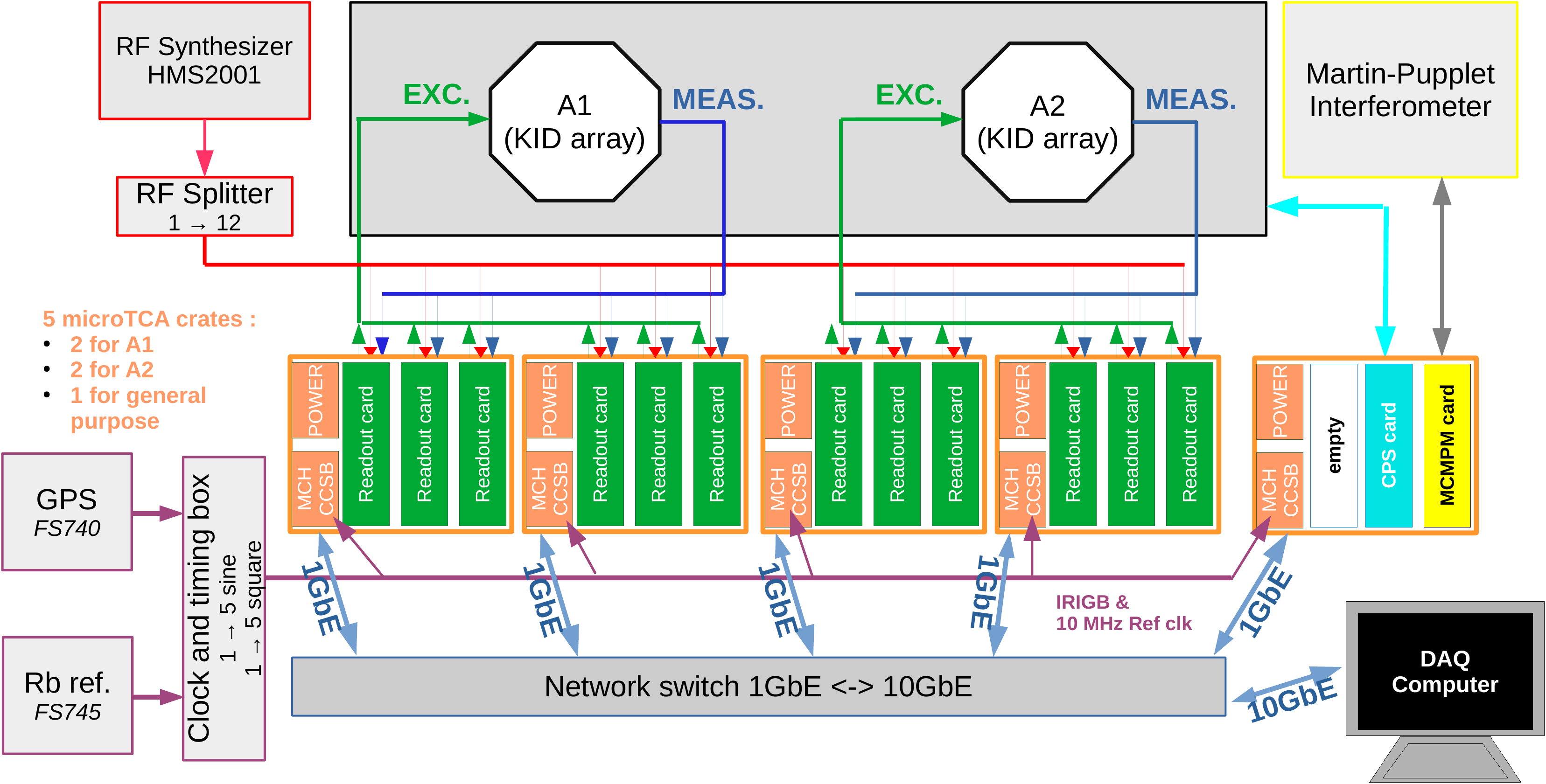}
\caption{\label{crateFig} Five microTCA crates are used to host most of the warm electronics (twelves readout boards and two monitoring and control boards). The timing and control box is in charge of distributing the 10\,MHz reference clock and the demodulated Inter-Range Instrumentation Group standard B (IRIG-B) signals to all crates. 
The Clocking and Timing Box distributes the reference clock and modulated IRIG-B provided by the reference.
The Data Acquisition (DAQ) computer is responsible for setting up the instrument at startup and then to record data during observations.
}
\end{figure}
As shown in figure~\ref{crateFig}, four crates are used to specifically host the KID readout electronics.
They are organized in two crates of three boards per array. 
Furthermore one crate is used to host the MCMPM and CPS boards.
Each crate features one central clocking and synchronization board (CCSB) mounted on a MicroTCA Carrier Hub (MCH) and one 600\,W power supply module.
The MCH, which is required by the microTCA specification, is in charge of managing the crate (slot activation for hot-plug, power supply monitoring, fan speed controlling, sensors monitoring). 
It hosts a Gigabit Ethernet (GbE) switch function for communicating with each slot and the CCSB.
The CCSB collects and distributes the reference clock and the demodulated Inter-Range Instrumentation Group standard B (IRIG-B) provided by the clock and timing box (CTB) via the crate backplane.

The CTB is composed of two different Printed Circuit Boards (PCB).
It fans-out the 10\,MHz reference clock provided by a FS725 Rubidium frequency standard (Stanford Research Systems) and the demodulated IRIG-B signal.
The latter is produced, before fan-out, by the CTB, which uses the modulated IRIG-B-AM  signal supplied by the GPS (FS740 from Stanford Research Systems).

A single RF synthesizer (Holzworth HSM2001A) is used to provide the Local Oscillator (LO) to each KID\_READOUT board, thanks to a commercial RF passive splitter.
The RF synthesizer uses the Rubidium 10\,MHz as a reference clock.

One modulation output from one of the KID\_READOUT boards is connected to the synthesizer in order to control the LO frequency which then permits to tune and calibrate the LEKIDs detectors.
The tuning is the process that allows to find the optimal excitation frequencies for each LEDKID \cite{bounmy2022}, while the calibration is used to determine the LEKID response to a known frequency shift.
All readout boards are operated with the same reference clock and started simultaneously using a Pulse Per Second (PPS) signal which is extracted from the IRIG-B frame received by the CCSB and then distributed to all AMC boards.

%%%%%%%%%%%%%%%%%%%%%%%%%%%%%%%%%%%%%%%%%%%%%%%%%%%%%%%%%%%%%%%%%%%%%%%%%%%%%%%
\section{KID readout board}
\subsection{Hardware description}
\begin{figure}[hbtp]
\centering
\includegraphics[angle=0,width=0.49\textwidth]{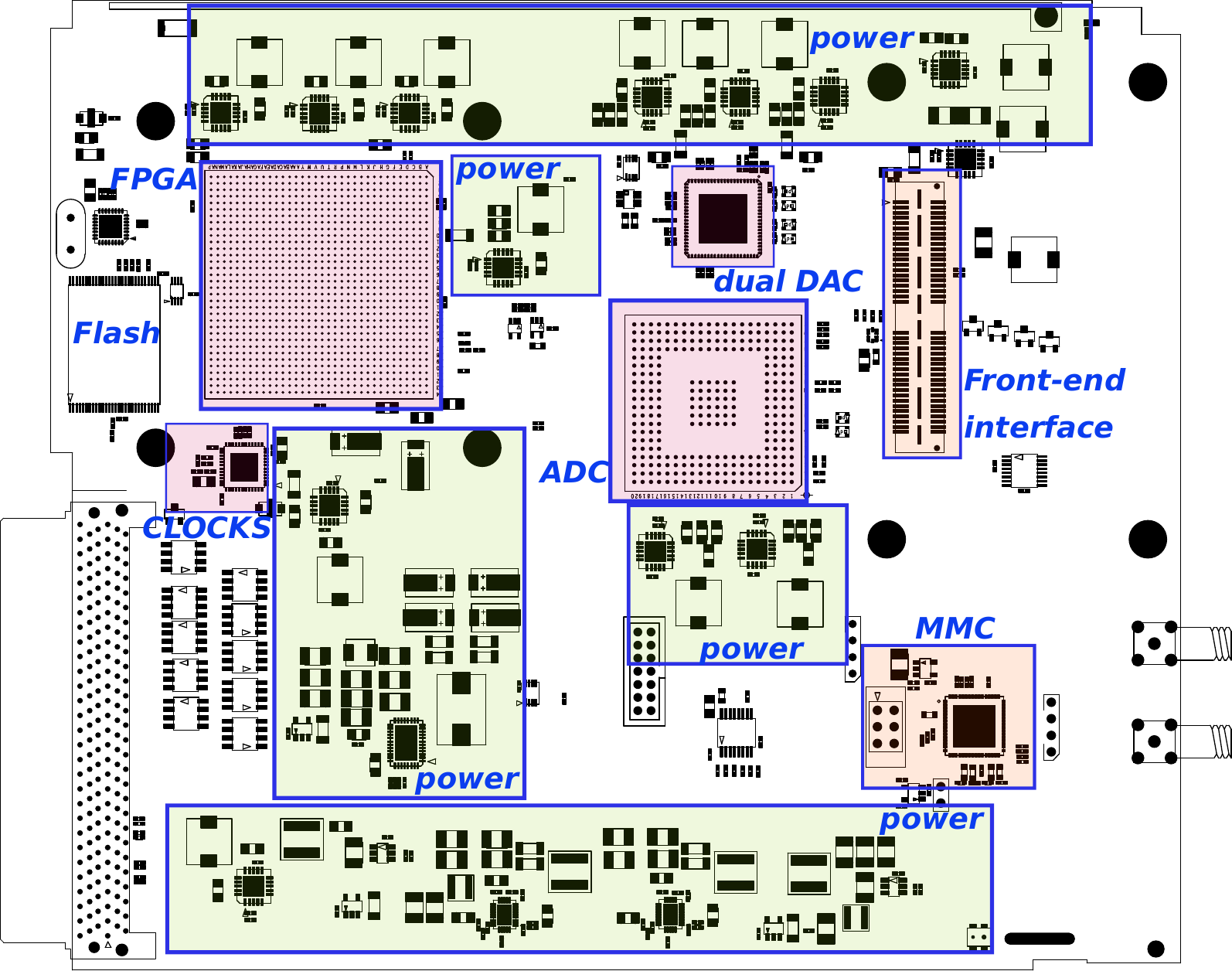}
\includegraphics[angle=0,width=0.49\textwidth]{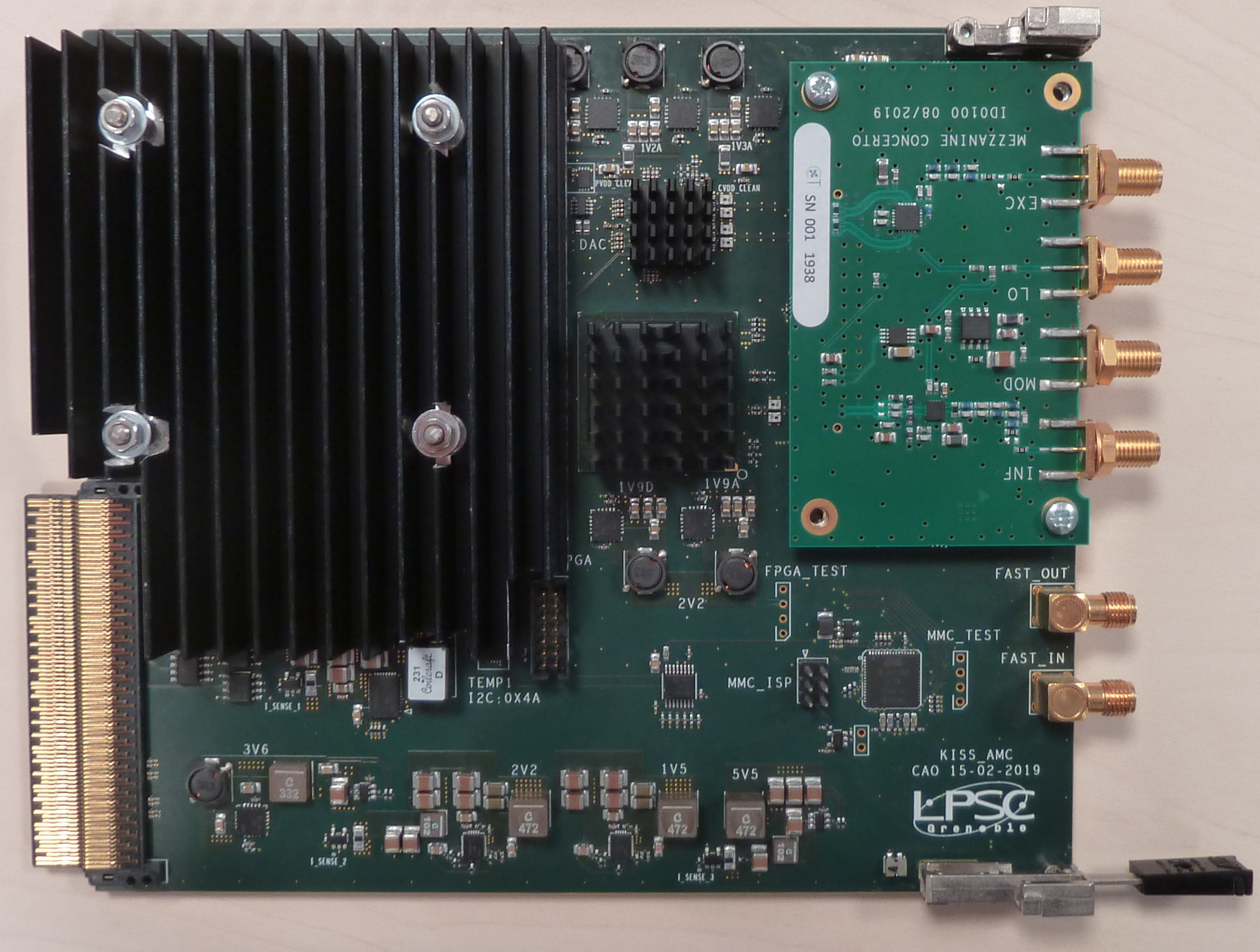}
\caption{\label{readoutHwFig} Overview of the KID\_READOUT board shown on the l.h.s, the main active components are shown (FPGA, A/D and D/A converters, interface to the radio-frequency front-end board and  power supplies). On the r.h.s, an actual board equipped with the heat sinks and the front-end board is shown.
}
\end{figure}

The KID\_READOUT board is composed of two parts, the digital and carrier part and the radio-frequency front-end part. 
The module fits in a full size double width AMC format (see figure~\ref{readoutHwFig}).
The carrier main components are a large Field Programmable Gate Array (FPGA) (Xilinx XCKU060FFVA1156-2), a 12-bit Analog to Digital Converter (ADC) (Texas Instrument ADC12D1000) and a dual 16-bit Digital to Analog converter (DAC) (Analog Devices AD9136). 
All converters are operated at 2\,GSamples/s.

The FPGA is the center part of the carrier board and it is connected to the ADC, the DAC and the crate back-panel.
The dual DAC and the ADC analog parts are connected directly to the front-end board via a dedicated high performance connector (SAMTEC \mbox{QSE-040-01-F-D-A}).
The FPGA generates the excitation comb digital signal sent to the DAC and analyzes the returning signal sampled by the ADC.
It is also in charge of interfacing the acquisition with the Gigabit Ethernet link though the MCH and the Clocking and Timing system via the CCSB.
The FPGA is configured at boot time from dedicated flash memory that hosts the firmware. 
This flash memory can be reprogrammed \emph{in-situ} through the Ethernet link, and thus allows remote firmware upgrades.

The main components are all clocked with a dedicated clocking circuitry which is referenced by the 10\,MHz clock distributed by the CCSB through the back-panel. Thus all readout boards are all perfectly synchronous and uses a very stable timing reference.

As shown in figure~\ref{readoutHwFig}, a large part of the carrier board is devoted to local power supply generation.
Indeed, 15 different voltages must be generated to separate analog from digital supplies and to optimize the power consumption: five for the FPGA, two for the ADC, five for the DAC, one for the RF part and two for the clocking part. 

A Module Management Controller (MMC) is implemented to comply with the microTCA standard. 
It is composed of an 8-bit micro-controller and hosts a dedicated firmware.
Aside from the mandatory functionalities required by the standard, it provides board health monitoring such as local power supplies current and voltage monitoring, board temperatures (four distributed LM35B probes) and power supply sequencing.
 
\begin{figure}[hbtp]
\centering
\includegraphics[angle=0,width=0.39\textwidth]{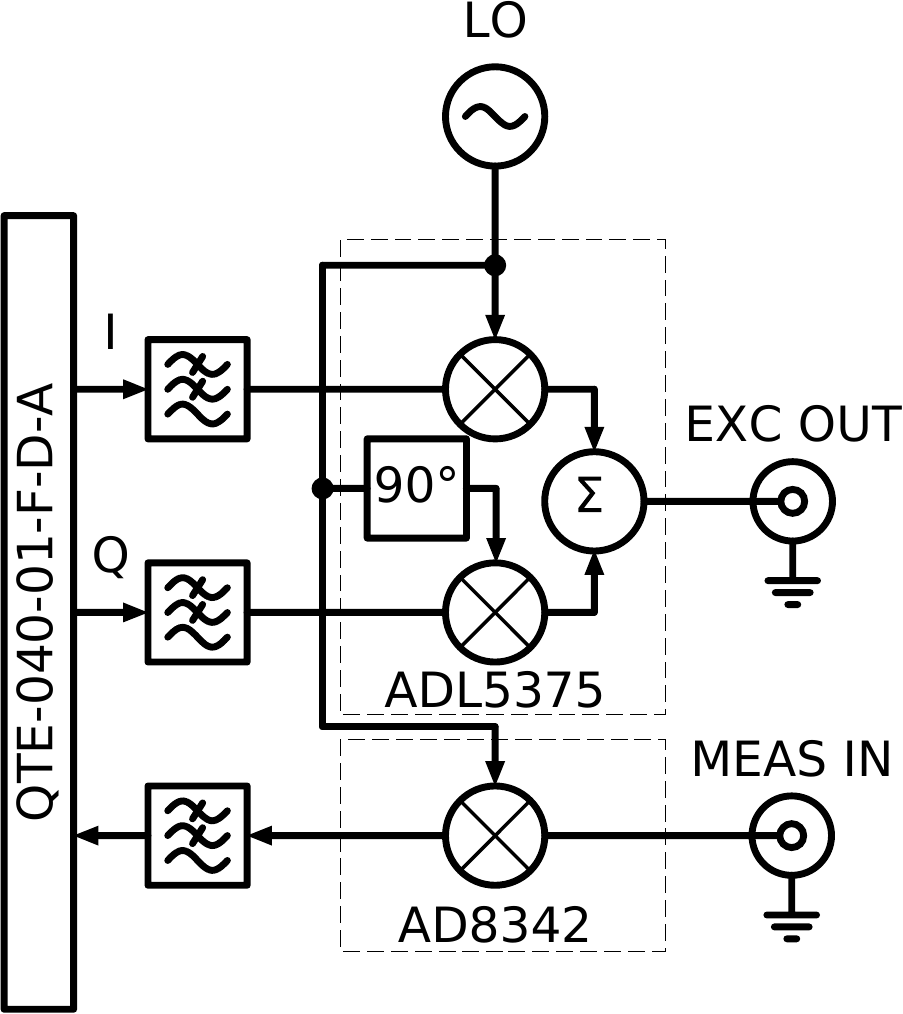}
\caption{\label{schMezzFig} 
Simplified schematic of the mezzanine board for signal shaping.
At first a low pass filter stage, composed of the SMC component LP0BA0790A7TR250 from AVX, suppresses the sampling frequency of the DAC and rejects the image response. 
The low pass filters have a cutoff frequency of about 950\,MHz.
The differential IQ signals filtered are mixed by an IQ mixer (ADL5375) that up-converts frequencies from DC-1\,GHz bandwidth to the LEKID matrix frequency range of 1.5\,GHz-2.5\,GHz, thanks to the Local Oscillator of 1.5\,GHz. 
Then, the up-converted signal (EXC\_OUT) goes through the cryostat, probes the state of the detectors (MEAS) and is down-converted at the input of the mezzanine thanks to a mixer (AD8342) with the same LO used previously. 
}
\end{figure}

Connected to the main board, a mezzanine contains the front-end analog shaping electronics, as shown in figure \ref{schMezzFig}. 
At first a low pass filter stage, composed of the SMC component LP0BA0790A7TR250 from AVX, suppresses the sampling frequency of the DAC and rejects the image response. 
The low pass filters have a cutoff frequency of about 950\,MHz.
The differential IQ signals filtered are mixed by an IQ mixer (ADL5375) that up-converts frequencies from DC-1\,GHz bandwidth to the LEKID matrix frequency range of 1.5\,GHz-2.5\,GHz, thanks to the Local Oscillator of 1.5\,GHz. 
Then, the up-converted signal (EXC\_OUT) goes through the cryostat, probes the state of the detectors (MEAS) and is down-converted at the input of the mezzanine thanks to a mixer (AD8342) with the same LO used previously. 
The resulting signal is then ready to be digitized after low-pass filtering.

\subsection{Firmware description}

\begin{figure}[hbtp]
\centering
\includegraphics[angle=0,width=0.99\textwidth]{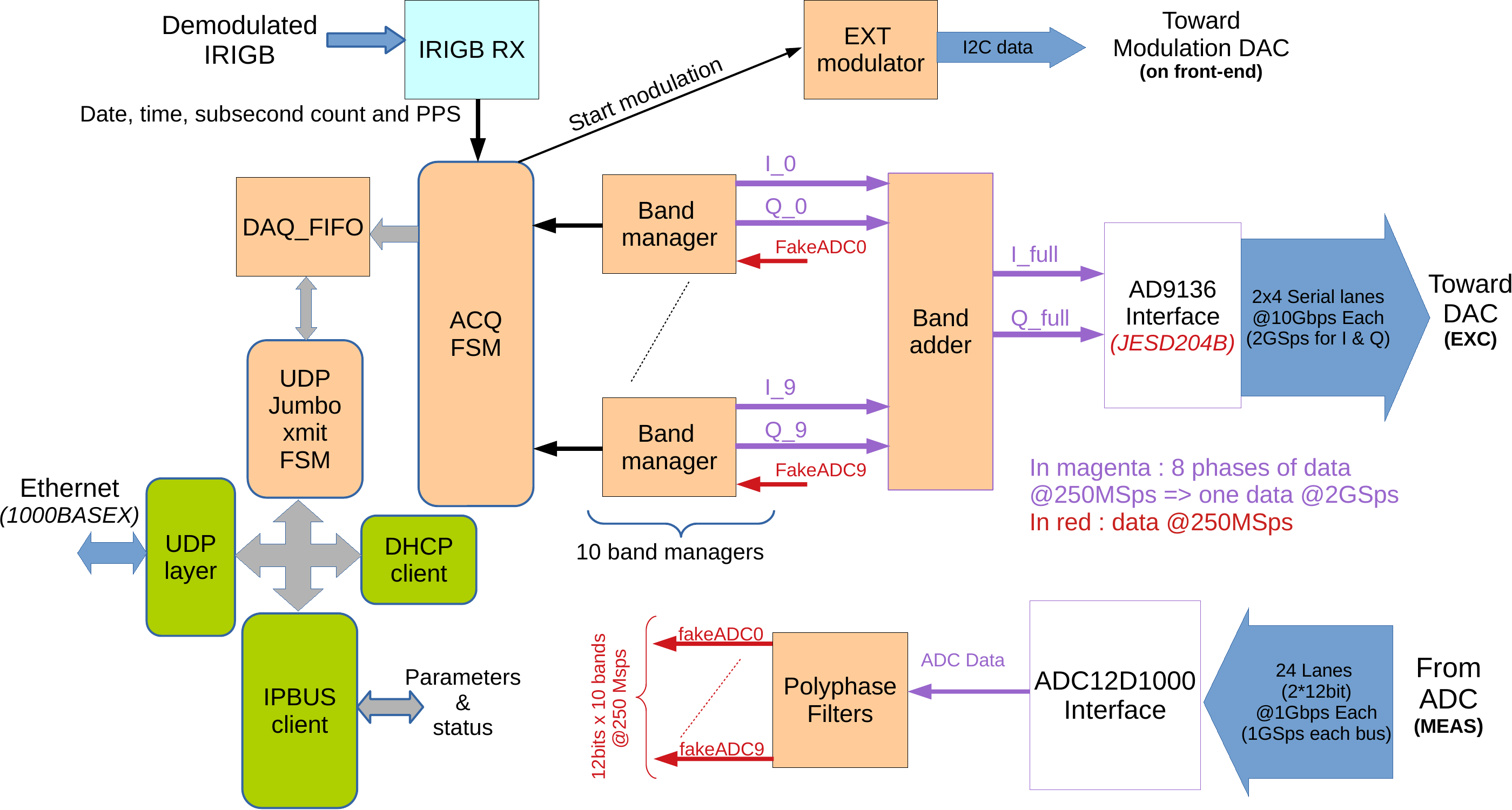}
\caption{\label{readoutFwFig} 
Overview of the FPGA firmware used for the KID readout.
}
\end{figure}
An overview of the firmware is shown in figure~\ref{readoutFwFig}. 
The center piece is the acquisition Finite State Machine (ACQ\_FSM) which collects the 400 I,Q data generated by the band managers and tags them with precise timing thanks to the IRIG-B receiver.
One data frame holds two 32-bit words (I,Q data) per LEKID resonator monitored (400 in total), the timing information on two 32-bit words, the frame number on one 32-bit word and eleven ADC/DAC monitoring 32-bit words (peak values observed). 
Thus, it is composed of 3256 bytes of data in total.
The data collected are stored in an buffer (DAQ\_FIFO), and waiting there to be shipped to the acquisition computer through Ethernet.

The IRIG-B receiver uses the demodulated signal and the reference clock provided through the microTCA backplane to extract the date and time from the serial frame and the Pulse Per Second pulse (PPS).
This timing functionality is extended with a sub-counter incremented by the 10\,MHz reference clock and reset every PPS, hence a theoretical timing resolution of 100\,ns can be reached.
In practice the resolution is limited to about 4\,\textmu s because of the analog demodulation technique used in the CTB. 
However this is sufficient to synchronize the data that are sampled at a frequency of about 4\,kHz.

To synchronize the CONCERTO readout and control electronics, the acquisition and data recording is armed by software and started on the following PPS received.
This results in having all boards embedded in the microTCA crates begin the data production simultaneously and have the same frame numbering.
Then, when the astronomical observation is about to start, the external modulation sequence (EXT\_modulator) is programmed to start at a given frame number, which is the same start number programmed in the MCMPM for initiating the movement of the mirror of the MPI.

The EXT\_modulator element is made of a FSM coupled to a memory table holding the modulation values to apply at each sampling step. 
The memory table is large enough to hold all values for the slowest MpI operating frequency, which is 1\,Hz, i.e. 4096 steps of 1/4\,kHz.
The modulation values are transferred via I2C protocol to the 10-bit DAC (AD5311) installed on the RF front-end board. 
Both the KID\_READOUT boards and MCMPM board possess this memory table so the modulation, which used for tuning \cite{bounmy2022} and scientific calibration \cite{modulationFasano}, does not interfere with the interferometry measurement. 
For example, among the 4096 steps, only the first 64 steps are modulation samples leaving the rest for observation purpose.
These modulated data samples can be used to adjust each tone's probing frequency and as a reference response of the readout electronics to a known frequency shift of the LEKIDs \cite{bounmy2022}.

The communication between the acquisition computer and the FPGA is achieved via Ethernet through a UDP layer.
On the one hand, this layer allows us to set and monitor the board via the IPBUS protocol \cite{Larrea_2015}, which is a secured communication channel that offers an easy parallel interface on the user side.
On the other hand, it is used to send a dedicated UDP datagram (jumbo frame) every time at least one full acquisition frame is stored in the DAQ\_FIFO.
The use of large datagram maximizes the throughput and reduces the software overhead required to manage the data taking.
The average data rate due to the two arrays readout is about 150\,MB/s.

\begin{figure}[hbtp]
\centering
\includegraphics[angle=0,width=0.99\textwidth]{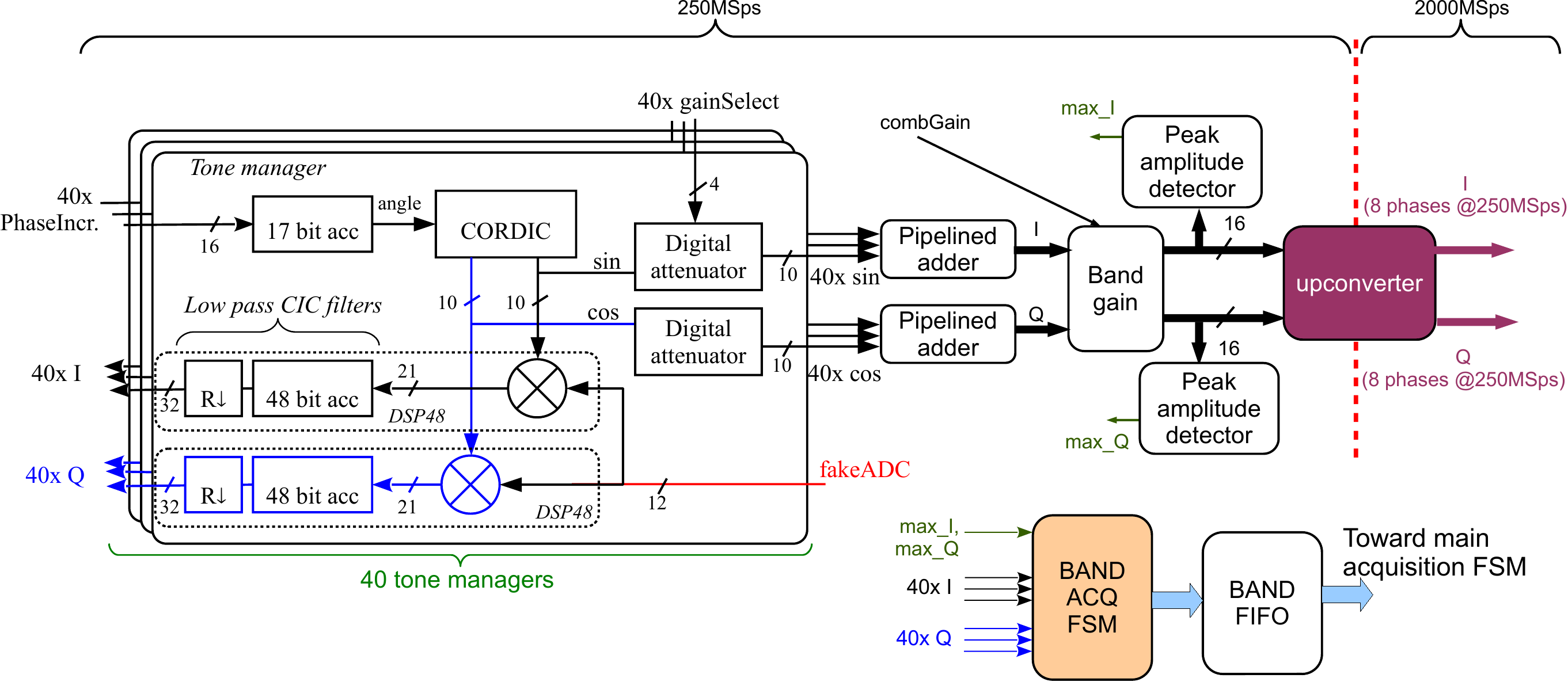}
\caption{\label{bandMgrFig} Band manager architecture. One band is composed of 40 tone managers that construct the excitation signals thanks to a CORDIC generator coupled to a phase accumulator and analyze the returning signal with a Digital Down Converter (DDC).
}
\end{figure}
To read-out the 400 LEKID resonators of a transmission line over a bandwidth of 1\,GHz, the firmware features ten band managers that are in charge of instrumenting a slice of 100\,MHz of the bandwidth.
The motivations for this configuration were to have a more optimal polyphase filters implementation and a lower operating frequency (250\,MHz) for the LEKID processors as it was the case in \cite{Bourrion_2016}.
As shown in figure~\ref{bandMgrFig}, each band manager contains 40 tone managers.
Each of these constructs the excitation signals thanks to a 10-bit CORDIC \cite{Volder1959TheCT} generator coupled to a phase accumulator and analyzes the returning signal with a Digital Down Converter (DDC) \cite{Lhning2000DigitalDC}.

The filters used in the DDC are Cascaded Integrated Comb filters whose decimation period is selected to be the same as one CORDIC phase accumulation ($2^{16}$ 250\,MHz clock cycles).
This yields an I,Q sampling period of about 4\,kHz.
The tone manager operates in the 250\,MHz clock domain and therefore the excitation signal must be up-converted to 2\,GSamples per second and frequency shifted in its 100\,MHz band, while the measurement signal must undergo the opposite processing.

On the excitation path, each sine and cosine signal are potentially attenuated with a digital attenuator by $\frac{1}{16^{th}}$ steps.
The 40 resulting sine and cosine signals are then numerically summed-up to construct the partial excitation comb which is then modified by a digital gain (16-bit tuning word).
The resulting signals are monitored by amplitude peak detector to ensure that no clipping occurs.
These peaks values (max\_I, max\_Q) are stored in the acquisition frame.
A clipping situation can be mitigated by either reloading the tone frequencies given the fact that each phase is quasi random and that an unlucky all-in-phase summation would then be removed, or alternatively by attenuating the gain for the concerned frequency band.

\begin{figure}[hbtp]
\centering
\includegraphics[angle=0,width=0.99\textwidth]{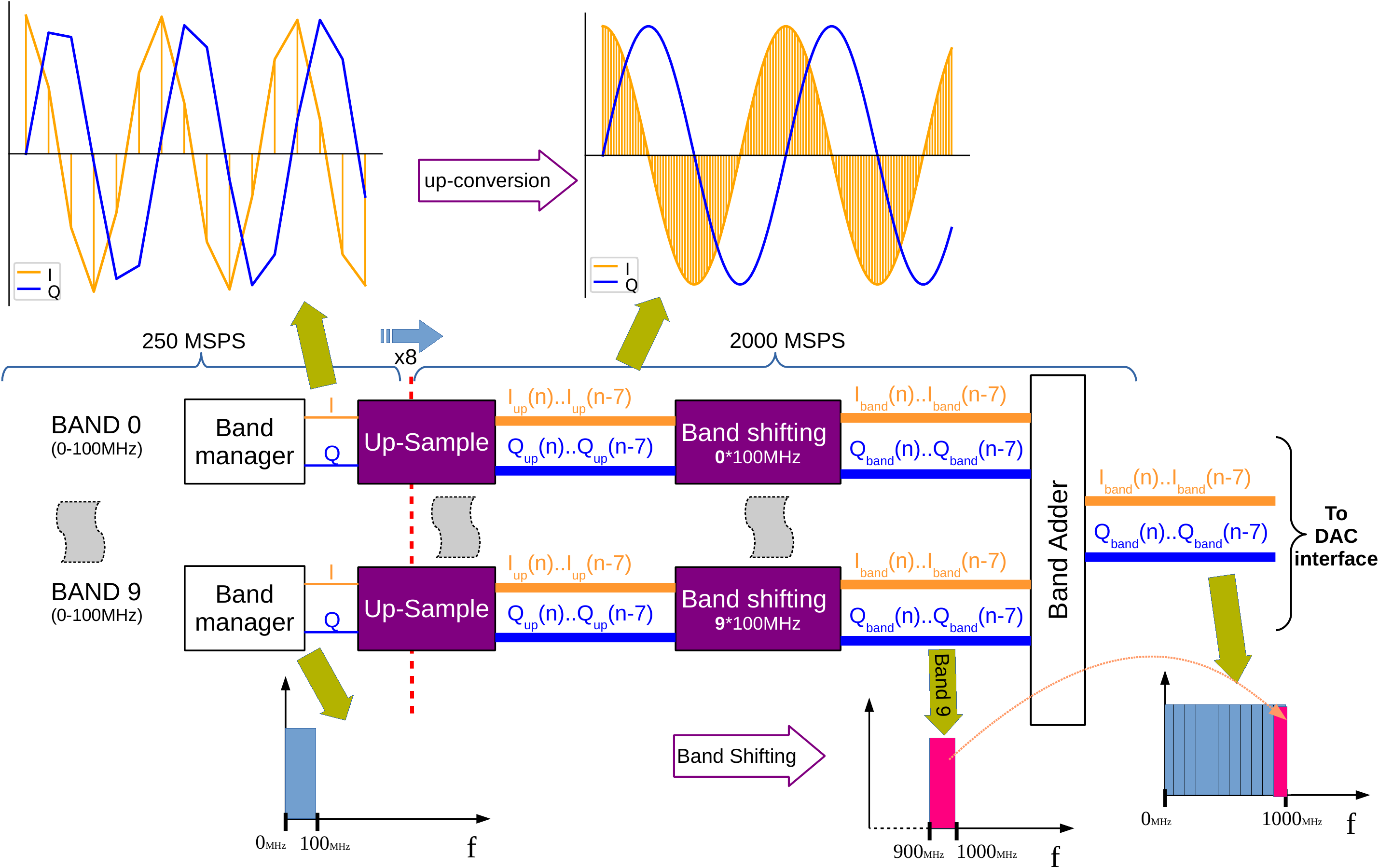}
\caption{\label{bandGenFig} Illustration of the excitation frequency comb construction. 
The full spectrum is composed of ten sub-bands having a width of 100\,MHz.
Each sub-band is generated at 250\,MSps and then up-converted at 2\,GSPs in its appropriate frequency band.
Each up-converter is composed of an up-sampler and a Numerically Controlled Oscillator (NCO) ensuring the appropriate frequency band-shifting.
}
\end{figure}

Finally, the up-conversion operation is performed in each band manager.
As depicted in figure~\ref{bandGenFig}, each sub-band is generated at 250\,MSps and then up-converted at 2\,GSPs in its appropriate frequency band.
Each up-converter is composed of an up-sampler and a Numerically Controlled Oscillator (NCO) ensuring the appropriate frequency band-shifting.
The ten band excitation signals are then numerically summed and fed to the dual DAC.

%\begin{figure}[hbtp]
%\centering
%\includegraphics[angle=0,width=0.6\textwidth]{polyphase_filters.pdf}
%\caption{\label{polyphaseFig} 
%The polyphase filter takes the eight phases of the measurement signal digitized at 2\,GSps, does the ten required frequency shifting and the decimation to down-convert the full spectrum in ten sub-spectrum having a bandwidth of 100\,MHz.
%}
%\end{figure}
The signal (fakeAdc 1,2,3, ... in Fig.~\ref{readoutFwFig} ) used on the measurement path by each bandManager is provided by the adequate output of the polyphase filter as shown in figure~\ref{readoutFwFig}.
This filter takes the eight phases of the measurement signal digitized at 2\,GSps, performs the ten required frequency shifting and the decimation to down-convert the full spectrum in ten equally spaced spectrum, each having a usable bandwidth of 100\,MHz.

\subsection{Resource usage}
A significant effort was made to optimize overall the firmware design for the experiment.
In particular, the polyphase filter and the up-converters were studied and designed to use the minimum number of DSP while providing adequate performances for the experiment. 
Consequently, the firmware uses about 214k (64.6\%) Configurable Logic Block (CLB) Look Up Table (LUT), 345k (52.1\%) CLB Flip-flops (FF)  and 1973 (71.5\%) Digital Signal Processors (DSP) of the available resources.
The detailed allocation of resources is shown in table~\ref{resourceTable}.

\begin{table}[hbtp] 
\centering
\includegraphics[angle=0,width=0.95\textwidth]{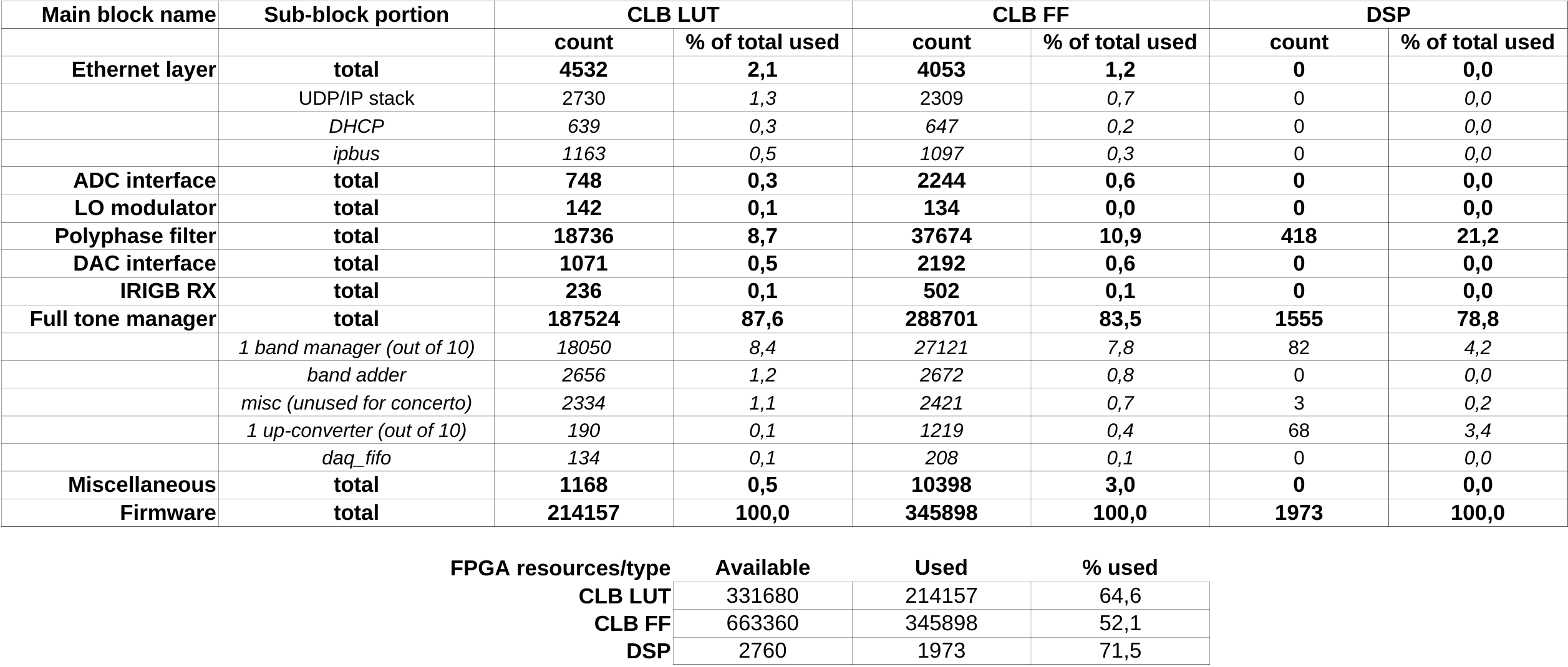}
\caption{Table summarizing the FPGA resource usage of the KID READOUT board.
The various firmware components and their relative contribution to the total are shown.
The overall FPGA resource consumption with respect to the maximum available is also given.
}
\label{resourceTable}
\end{table}
From the table, it can be seen that the highest resource users are the band managers, each of them uses about 8\% of the CLB resources and about 4\% of the DSPs.
In terms of DSP resources usage, the first users are the ten up-converters, for a total of 680 DSPs, and then the polyphase filter uses 418 DSPs.

%%%%%%%%%%%%%%%%%%%%%%%%%%%%%%%%%%%%%%%%%%%%%%%%%%%%%%%%%%%%%%%%%%%%%%%%%%%%%%%
\subsection{Readout electronics performances}
\subsubsection{Power consumption}
One KID\_readout board uses about 37\,W in total power to instrument 400 KIDs over the total bandwidth of 1\,GHz.
Four crates, each equipped with three readout boards, are required to readout the two arrays of 2152 pixels.
Within each crate, the power consumption of one MCH+CCSB is about 12\,W and each fan tray uses 17\,W.
Hence, each equipped crate uses 157\,W of power from the 12\,V back-plane distributed supply.
Accounting for the power module efficiency, about 90\%, each crate requires about 175\,W for the main power supply.
In the end the total power required for the four readout crate is about 700\,W below 0.17 W per KID.

\begin{figure}[hbtp]
\centering
\includegraphics[angle=0,width=0.9\textwidth]{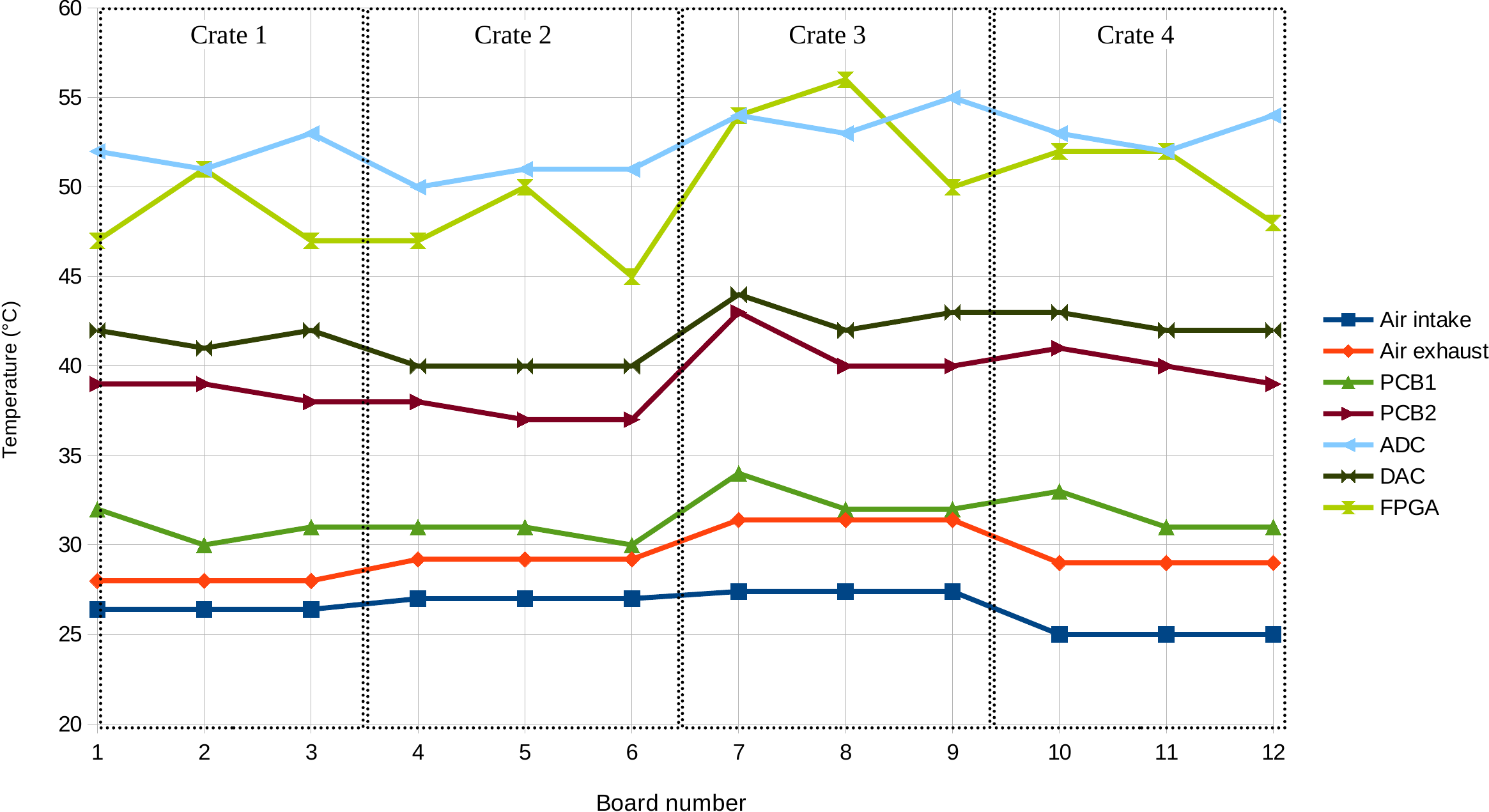}
\caption{\label{Temperatures} 
Temperatures plot as a function of the board number.
For each KID readout board, two PCB temperatures and FPGA, DAC and ADC die temperatures are shown.
MicroTCA air intake and exhaust temperatures for each crate are also shown.
A maximum temperature elevation of about 29\,°C is observed for the FPGA and ADC components.
}
\end{figure}

To ensure safe operation, various temperatures are constantly monitored on the experimental site, thanks to the features provided by the KID readout boards and the microTCA crates.
Indeed, each KID readout board is equipped with sensors probing two PCB temperatures and three others for the FPGA, DAC and ADC die temperatures, while the microTCA crates measure the air intake and exhaust temperatures. 
The plot of these temperatures is shown in figure~\ref{Temperatures} as a function of the board number.
We can see that the maximum temperature elevation is about 29\,°C for the FPGA and ADC components.

\subsubsection{System transfer function}

\begin{figure}[hbtp]
\centering
\includegraphics[angle=0,width=0.49\textwidth]{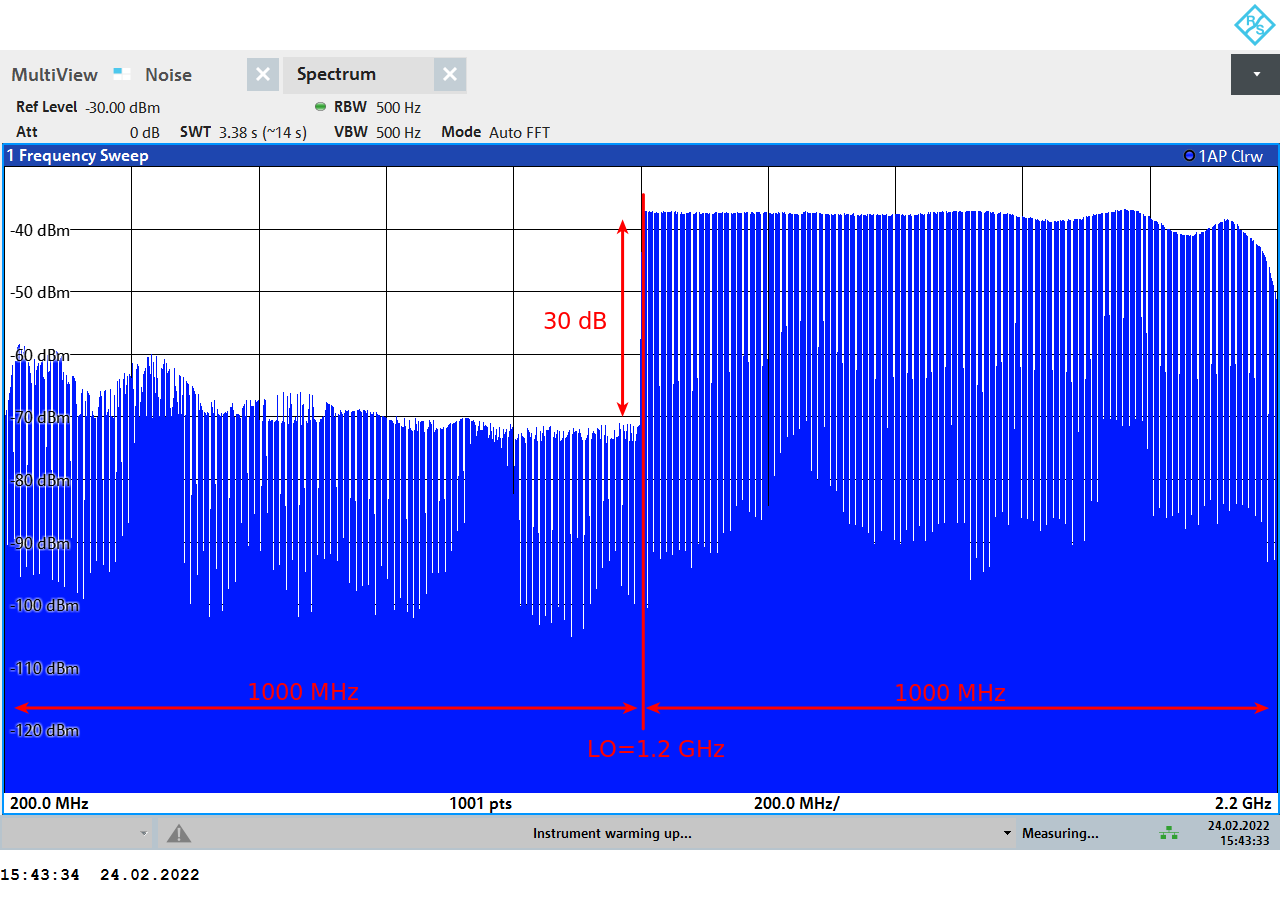}
\includegraphics[angle=0,width=0.49\textwidth]{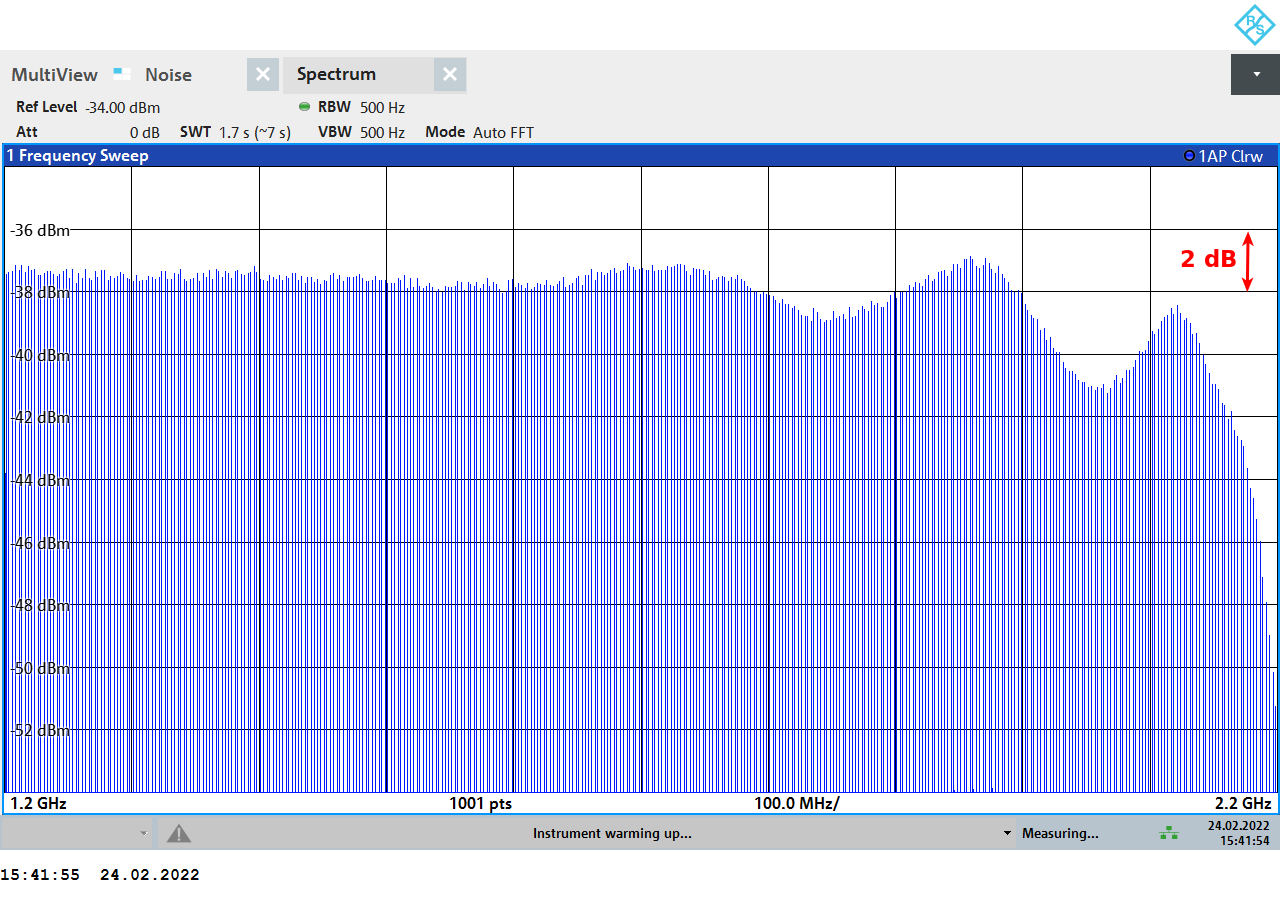}
\caption{\label{spectrumSnapshots} 
Excitation output spectrum measurement with Local Oscillator (LO) at 1.2\,GHz.
The spectrum on the l.h.s was taken with a frequency span of 2\,GHz, while the one on the r.h.s. was taken with a frequency span of 1\,GHz and with dilated amplitude scale.
The flatness of the frequency response can be seen, as well as the steep analog low pass filter effect induced at base band (DAC output, ADC input). 
With up and down conversion, this translates to steep drop from LO+950\,MHz and above.
}
\end{figure}

As a first characterisation of the electronics, an excitation signal composed of 400 tones evenly spaced in frequency was generated by the KID\_readout board and measured with a Vector Network Analyzer (Rhode \& Schwarz ZNL6).
The Local Oscillator (LO) input was set at 1.2\,GHz, the measured spectra can be seen in figure~\ref{spectrumSnapshots}.
The nice side band rejection (-30\,dB), as well as the flatness (<2\,dB) of the spectrum can be observed.

\begin{figure}[hbtp]
\centering
\includegraphics[angle=0,width=0.49\textwidth]{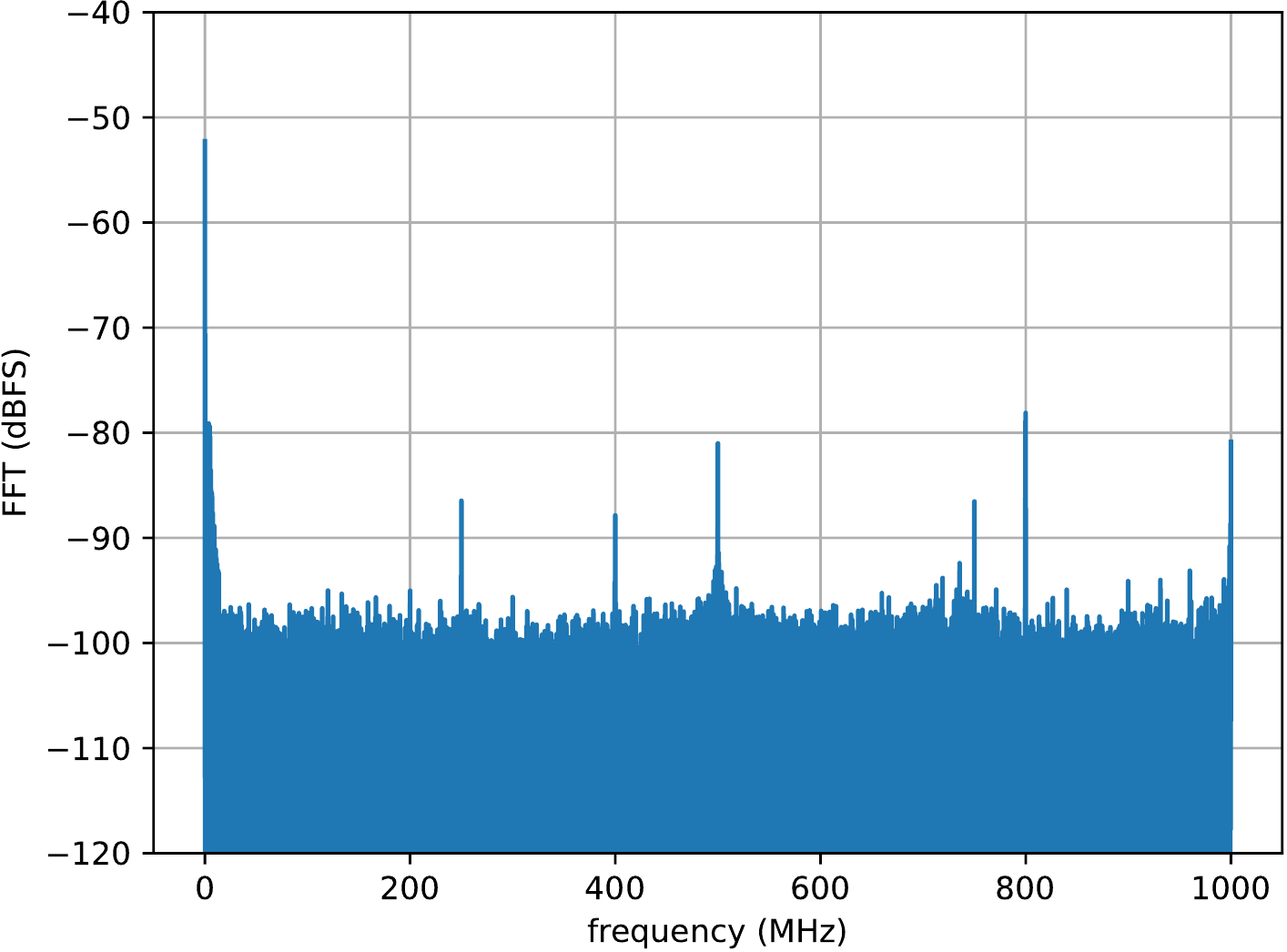}
\includegraphics[angle=0,width=0.49\textwidth]{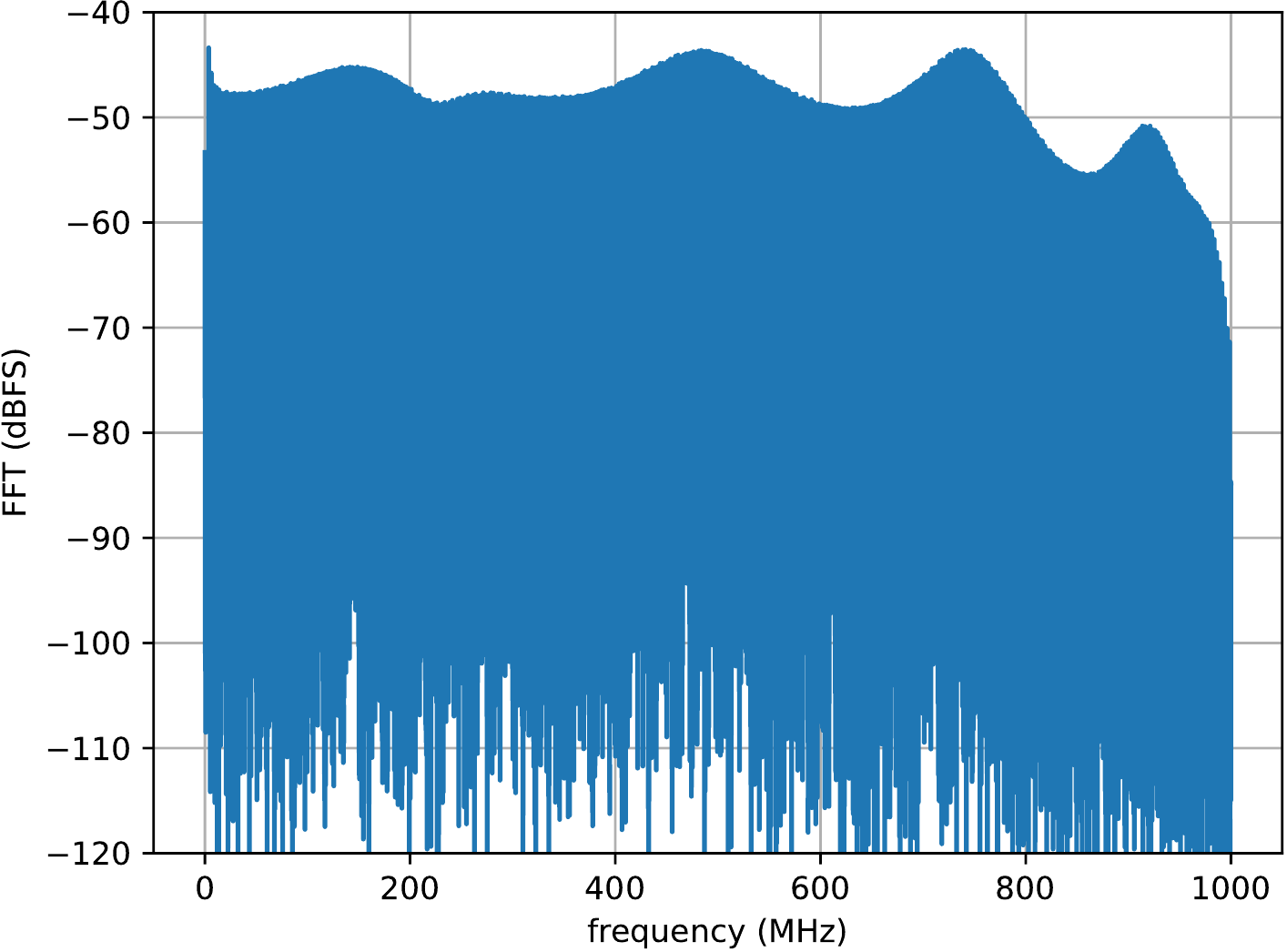}
\caption{\label{fftADCplots}
FFT plots of the ADC data recorded at 2\,Gsamples/s when operating in external loop-back mode.
The spectrum of the ADC response when no excitation signal is produced is shown on the l.h.s.
On the r.h.s. is shown the ADC response spectrum when an excitation signal composed 400 tones evenly spaced in frequency is generated.
}
\end{figure}

In a second stage, using the same LO frequency, the excitation output of the readout board was looped-back in the measurement input with an external cable.
Then using a direct ADC signal recording feature embedded in the FPGA firmware, two snapshots of the ADC output were recorded at 2\,Gsamples/s, each featuring 65356 samples.
The first snapshot was recorded with no excitation signal produced, while the second featured the 400 tones evenly spaced in frequency like previously.
For each snapshot, a Fast Fourier Transform (FFT) was computed (see figure~\ref{fftADCplots}).
On the spectrum of the ADC response with no excitation signal produced, we can see the harmonics of the 250\,MHz (reference clock used by the board), the 500\,MHz (ADC dual data rate clock frequency) and two peaks at 400\,MHz and 800\,MHz due to the inter-modulation products between the LO and the sampling frequency.
The spectrum of ADC response featuring the excitation signal is highly similar to the one recorded directly at the excitation output (see figure~\ref{spectrumSnapshots}), aside from the end of the bandwidth (above 950\,MHz) where the drop in response is much more important.
This can be explained by the additional low pass filter embedded on the mezzanine board (figure~\ref{schMezzFig}) used to filter the ADC analog input.

\subsubsection{System noise}
\begin{figure}[hbtp]
\centering
\includegraphics[angle=0,width=0.99\textwidth]{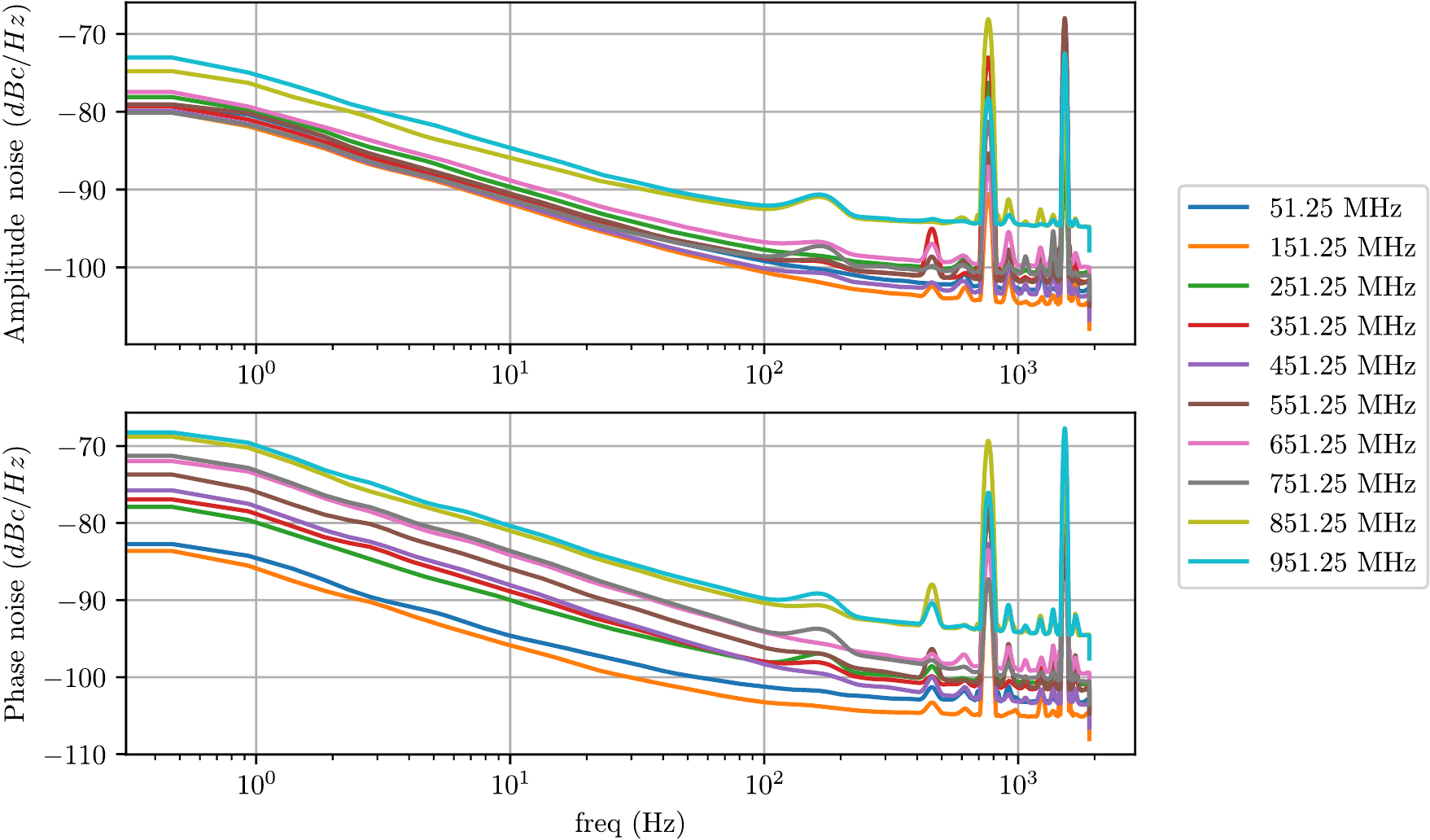}
\caption{\label{PSDspectrum} 
Power spectral density plots showing amplitude (top) and phase noise (bottom) for one tone in each band. 
The Welch method was used to compute the noise power spectra.
The window lengths were adapted to the frequency interval (smaller/larger window lengths at high/low frequency) to smooth the power spectra.
}
\end{figure}

\begin{figure}[hbtp]
\centering
\includegraphics[angle=0,width=0.99\textwidth]{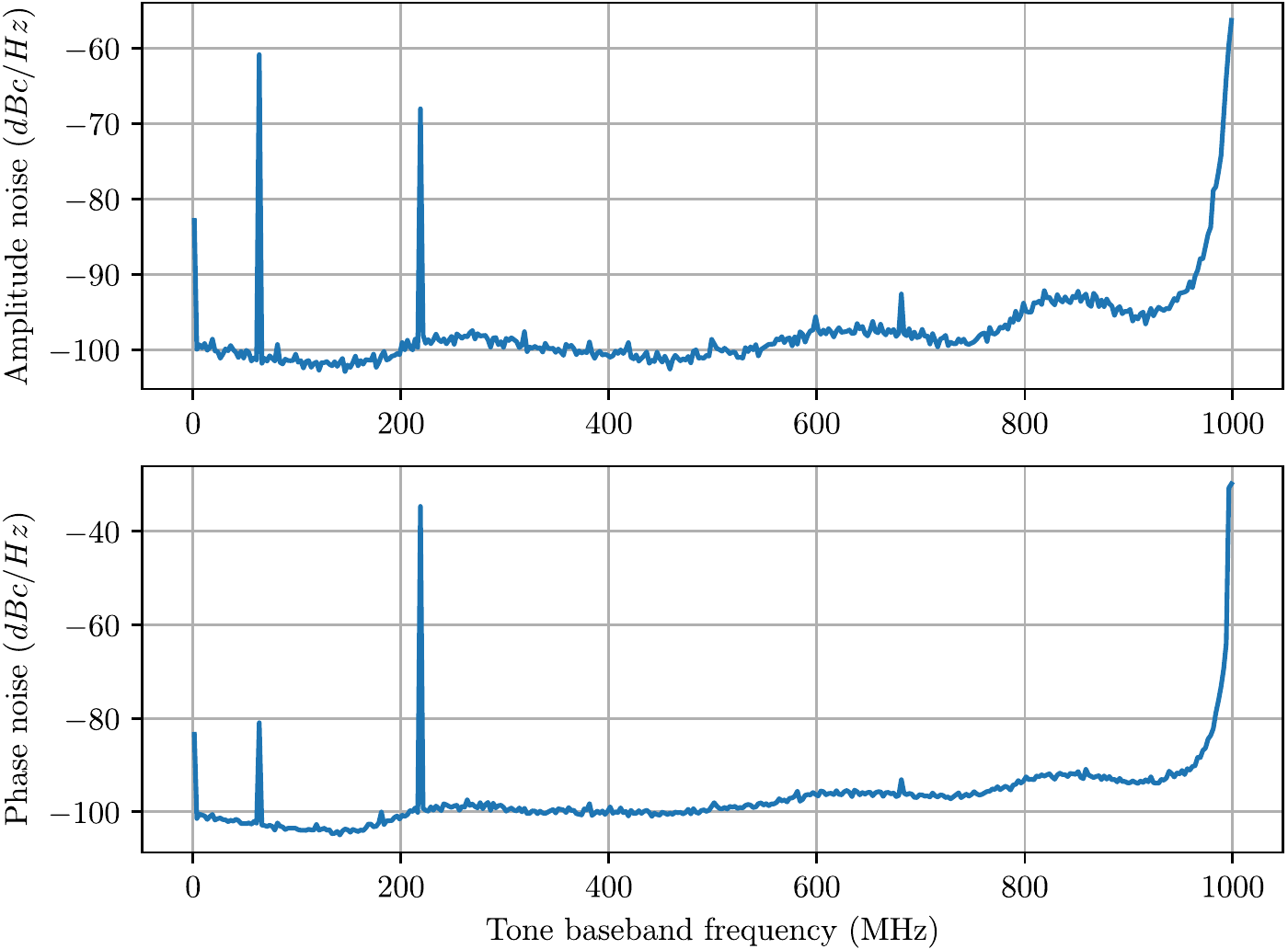}
\caption{\label{noiseDist} 
Average noise per tone between 110\,Hz and 280\,Hz. The 400 tones are evenly distributed over the 1000\,MHz bandwidth.
}
\end{figure}

To assess the system noise the same frequency comb featuring 400 tones evenly spaced in frequency was used and 655360 IQ data were recorded for each tone.
As previously, the excitation output of the readout board was looped-back in the measurement input with an external cable and the LO was set at 1.2\,GHz.
The power spectral density of amplitudes and phases were computed for one tone in each band (see figure~\ref{PSDspectrum}).
These plots show that the system noise floor is reached at around 100\,Hz and that it is in the order of -100\,dBc/Hz.
We also observe two unwanted peaks at 763\,Hz and 1527\,Hz.
Unfortunately, those are not yet explained, but they are sufficiently far from the scientific signal, which lies in the 110\,Hz to 280\,Hz range.
Consequently, to have an overview of the phase and amplitude system noise in this area of interest for the full electronics bandwidth, the average noise of each tone between 110\,Hz and 280\,Hz is plotted in figure~\ref{noiseDist}.
We can see that the noise level stays around -100\,dBc/Hz over the full bandwidth and experience a significant rise in the end of the spectrum (after 950\,MHz), this is consistent with the fact that the transmission response drops in this area (see figure~\ref{fftADCplots}).
The peaks observed at 62.5\,MHz and 250\,MHz are respectively sub-products of the Ethernet reference clock (125\,MHz) and the ADC and DAC reference clock (250\,MHz).

%%%%%%%%%%%%%%%%%%%%%%%%%%%%%%%%%%%%%%%%%%%%%%%%%%%%%%%%%%%%%%%%%%%%%%%%%%%%%%%
\section{Motor Controller and Martin-Puplett Monitor (MCMPM) board}
\begin{figure}[hbtp]
\centering
\includegraphics[angle=0,width=0.8\textwidth]{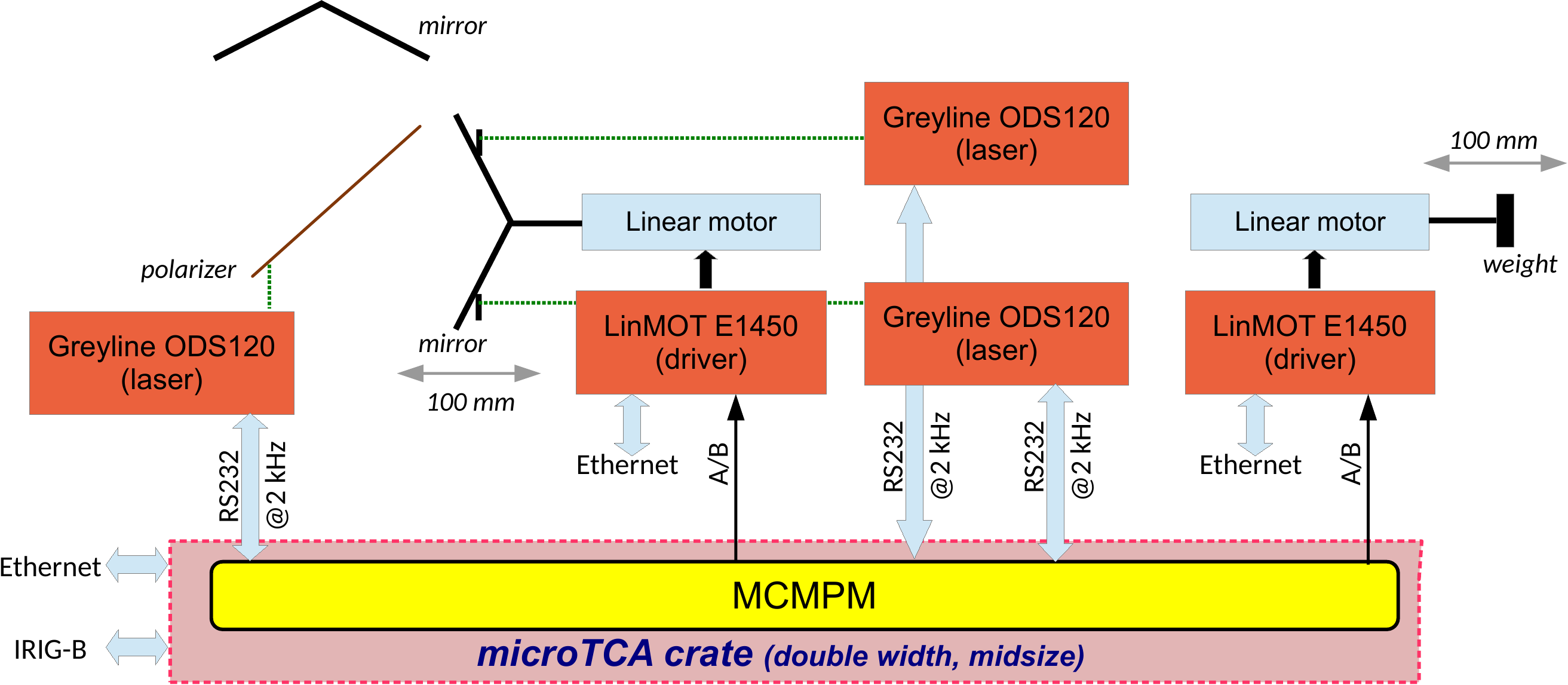}
\caption{\label{mcmpmGenFig} 
Overview of the Martin Puplett Interferometer control electronics. 
The board is inserted in the microTCA crate and communicate via IPBUS with the DAQ software.
On the front-end side, it is connected to three position measuring devices via serial links and to two linear motor drivers.
}
\end{figure}

The Martin-Puplett Interferometer linear motors drivers and mirror position monitors are managed by a single dedicated board called the Motor Controller and Martin-Puplett Monitor (MCMPM) board (see figure~\ref{mcmpmGenFig}).
The two main requirements of this board are (i) to always move simultaneously and in opposite direction the mirror and its counter-weight to dampen the vibrations to a minimum and (ii) to drive the motors and monitor the positions synchronously with respect to the KID readout electronics.

The board was designed to be hosted in a microTCA crate. 
On the back-end side, through the back-panel connector, it communicates via Ethernet (Ipbus) with the DAQ software and receives the 10\,MHz reference clock and the PPS signal.
On the front-end side it is connected to two linear motor drivers and to three position measuring devices via serial links.
Two of these are used to measure the mirror position and its possible deformations, while the third is used to monitor vibrations of the polariser membrane, which changes the effective optical path difference \cite{monfardini2021concerto}.
The interface between the back-end and front-end is ensured by an XC7K70TFB676-1 Xilinx FPGA.

To avoid parasitic signals and unwanted smearing in the acquired interferograms and hence ease the data analysis, the motor movements and various position measurements must be done synchronously to the data acquisition \cite{fasano_spie}.
Consequently, the two stepper driver (linMOT E1450) are driven with an incremental code (A/B) generated by the MCMPM.
At the same time, the mirror top and bottom positions are measured with LASER (Greyline ODS 120) measuring devices that are triggered and read-out by the MCMPM via a communication serial link (RS232 protocol).
The linear motor driver parameters are configured through Ethernet with a driving software.

\begin{figure}[hbtp]
\centering
\includegraphics[angle=0,width=0.9\textwidth]{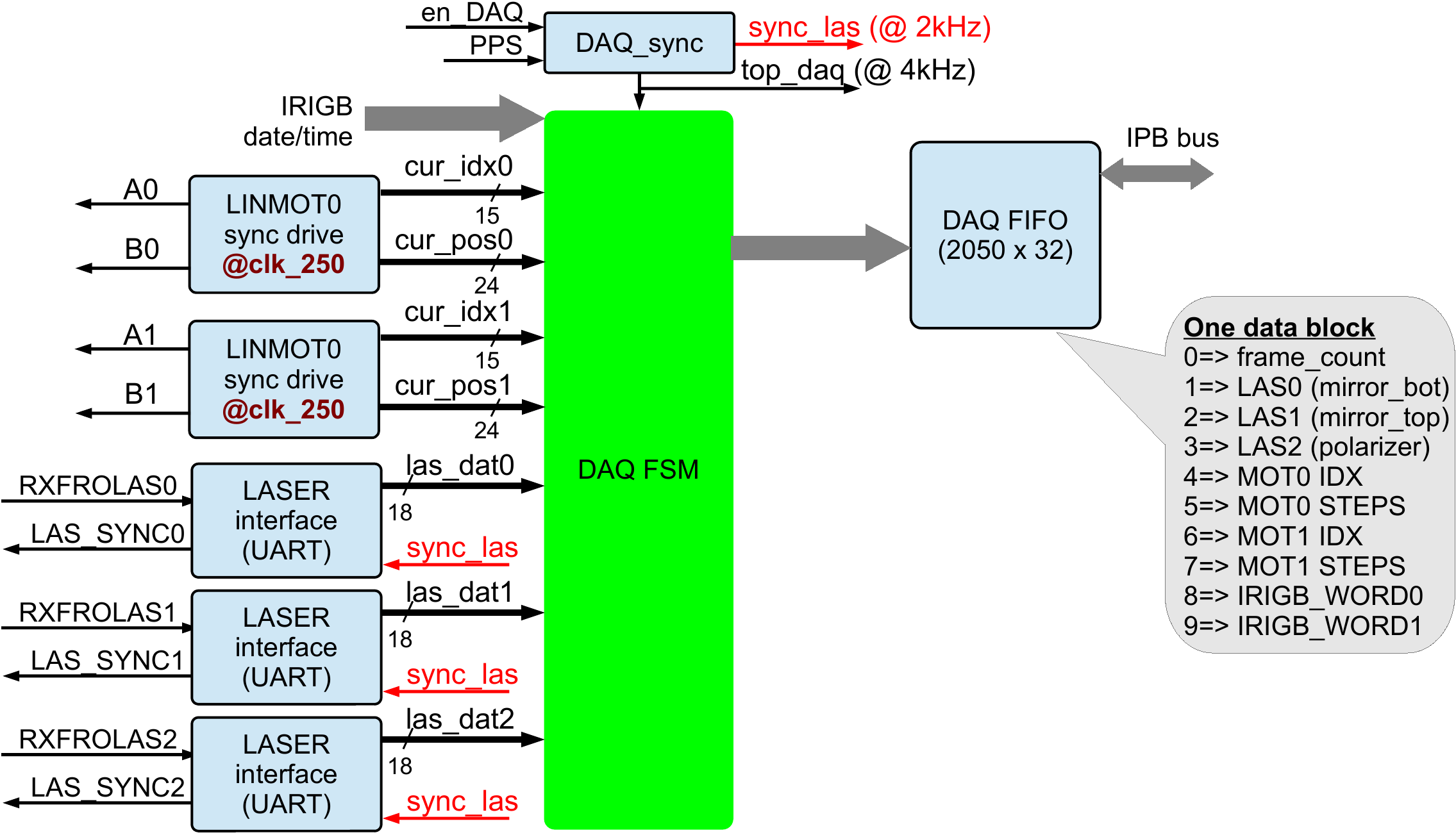}
\caption{\label{mcmpmFwOverviewFig} 
MCMPM firmware overview. 
Interfaces with the three position monitoring devices and the two linear motor drivers are shown. 
The FIFO interface, as well as one data frame format are shown on the r.h.s.
}
\end{figure}
An overview of the firmware is shown in figure~\ref{mcmpmFwOverviewFig}.
On the right side, the interfaces with the position monitoring devices and the linear motor driver are shown.
In the center, sits the central Finite State Machine (FSM), which is in charge of synchronously acquiring the data.
On the left side, the buffer used to smooth the data flow sent toward the Ethernet (IPBus) and its content are shown.

Thanks to the 10\,MHz reference clock, the DAQ\_sync block produces the enable signal used to perform the acquisition cycles at the same rate as the KID data readout, i.e. at $\rm \dfrac{250}{2^{16}}\sim4\,kHz$.
However, the position measuring devices are limited to a maximum acquisition rate of 2\,kHz, so the LASER\_interface are triggered at half the acquisition rate.
Consequently, to up-sample the laser positions, each position measured is duplicated in two consecutive data blocks as those are generated at about 4\,kHz in the firmware.

The acquisition process, and hence the frame numbering, is started on the subsequent PPS received after the electronics is armed by software.
Using the same methodology for all electronic boards involved in the acquisition provides a simple way to start synchronously the acquisition from a GPS receiver.
In parallel, the decoded IRIG-B date and time code along with the sub-second counter are inserted in the data block by the DAQ FSM. 
This measurement, along with the frame number, permits (i) to uniquely time-tag each data block, but also (ii) to check the delay between the various readout electronics which must remain constant.
Indeed, the same IRIG-B signal is distributed on the five crates, these precise timing information allow the detection of an abnormal condition or drift in the DAQ and control system.

\begin{figure}[hbtp]
\centering
\includegraphics[angle=0,width=0.9\textwidth]{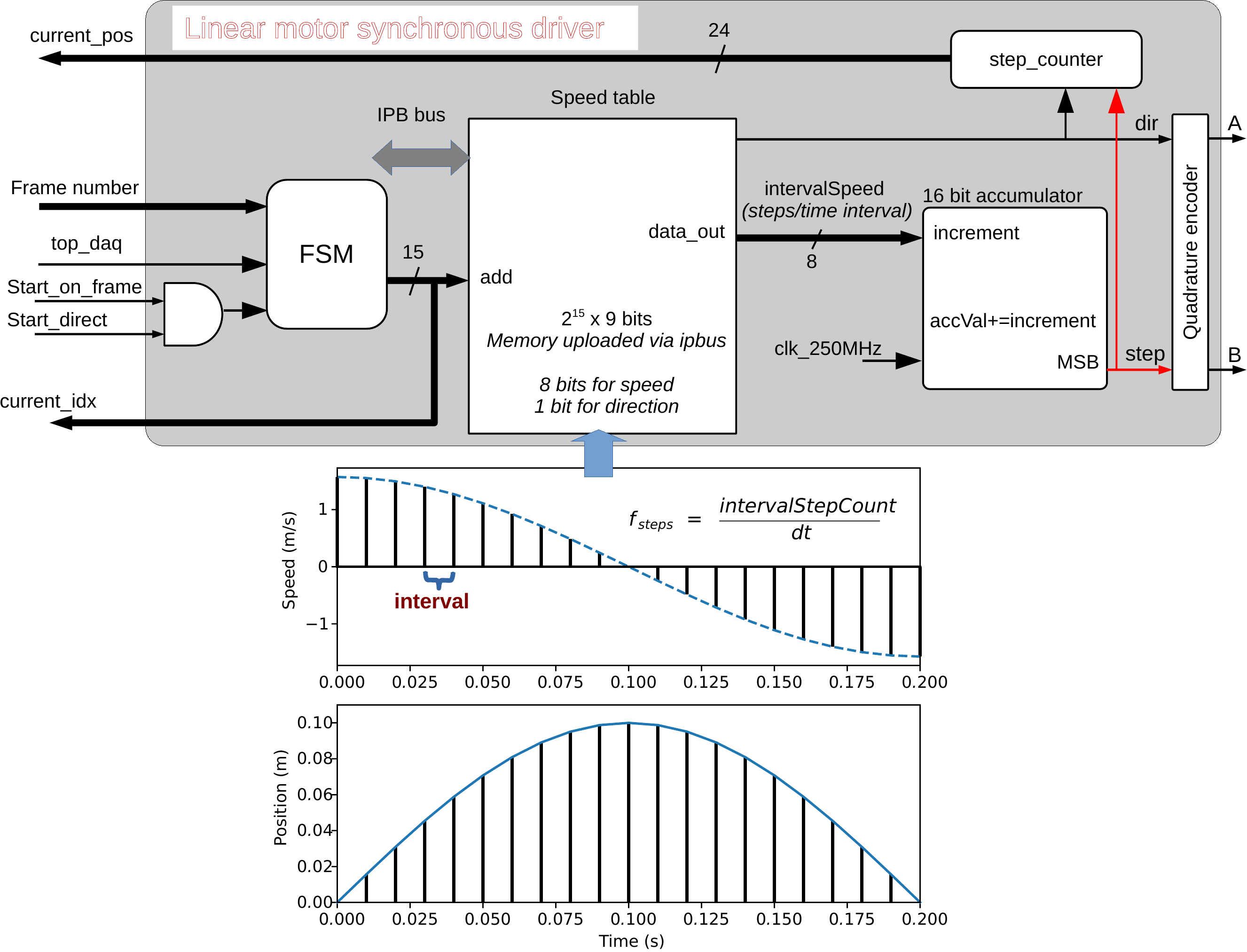}
\caption{\label{mcmpmSyncDriveFig} 
Block diagram of the linear motor controller. 
The core component of the controller is a large memory (32768 elements) that holds the steps frequency to generate along with the direction of the movement.
The motor movements are started upon the fulfilment of start conditions asserted by a dedicated Finite State Machine (FSM).
}
\end{figure}
The linear motor controllers (LINMOT0/1) are further detailed in figure~\ref{mcmpmSyncDriveFig}.
The core component of the controller is a large memory (32768 elements) that holds the steps frequency to generate along with the direction of the movement.
The memory depth was chosen in order to allow slow mirror frequency, i.e as low as 0.1\,Hz.
As illustrated in the two plots below the block diagram shown in figure~\ref{mcmpmSyncDriveFig}, the motor speed as a function of time can be derived from the mirror position as a function of the time.
The number of steps to produce per time interval is computed using the linear motor conversion factors, with the time interval being the duration of one acquisition cycle.
Finally, the motor step curve profiles are computed in advance and loaded in the MCMPM.
Given the fact that the mirror and its counter-weight must be moved simultaneously and in opposite direction to dampen the vibrations to a minimum, the curves are computed to be symmetric. 
The curve computing also uses the optical zero path difference offset and the maximum achievable acceleration (limitation) as input parameters.

At each time interval, the frequency word (named increment) provided by the memory is used by a 16-bit accumulator operating at 250\,MHz whose Most Significant Bit (MSB) is used to generate the steps needed by the motor power drivers.
The step frequency is given by the following formula: $\rm f_{step}=\dfrac{250\,MHz}{2^{16}}\times \dfrac{increment}{2}$.
Per manufacturer specification, the motor power drivers can accept a maximum step frequency of 2\,MHz, the increment width is limited to 8-bit, which 
allows the step frequency to be tuned between 0\,Hz and 1953\,kHz with a resolution of 7.6\,kHz.

The memory readout pointer is driven by a Finite State Machine, which starts its operation when armed by software and upon receipt of a frame number matching the configured start number.
This latter synchronization is required as the acquisition must be already running before starting mirror movements and we want the interferometer cycle to start just after the 3-point frequency modulation executed at each start of scan. 
Indeed this frequency modulation is mandatory before each interferogram acquisition as it allows to calibrate the KIDs response to a known frequency shift \cite{modulationFasano}.
The frequency modulation is injected by the KID readout board which is synchronized with the same technique as described here.

The firmware described above use a low percentage of the FPGA available resources, it uses 6805 (16.6\%) LUTs, 6953 (8.5\%) FFs and 36 (26\%) block RAMs.
The latter are mostly used for the IPBUS protocol as multi-buffering is required to allow lost packets recovery (16 block RAMs) and for the two speed tables (9 block RAM each).

%%%%%%%%%%%%%%%%%%%%%%%%%%%%%%%%%%%%%%%%%%%%%%%%%%%%%%%%%%%%%%%%%%%%%%%%%%%%%%%
\section{Conclusion}
In this paper we have presented a dedicated KID readout and control electronics for the CONCERTO experiment installed at the APEX 12-m telescope.
These electronics ensure the KID readout and the synchronous operation and monitoring of the CONCERTO MPI.
It consists of a total of five microTCA crates: four crates (three boards each) for KID readout and one crate for control and monitoring of the instrument.
With respect to previous KID readout electronics used in the NIKA and NIKA2 experiments it represents a major step forward.
It ensures similar noise properties with a much larger operation sampling rate (4\,kHz instead of tens of Hz), a doubled bandwidth (1\,GHz instead of 500\,MHz) and a more compact design with lower power consumption (35\,W per readout board).
Furthermore, by using a single DAC, it provides an homogeneous readout of the 400 tones avoiding the sub-bands uncorrelated noise contribution observed in the previous version featuring five DACs \cite{adam_phd}.
In addition, it allows us to synchronously and  accurately control the MPI motors, as well as monitor the position of the MPI mirrors and the vibrations of the MPI main polariser, that are critical for the scientific measurements.
Finally, these electronics have been successfully operated in real conditions (regular scientific CONCERTO observations are carried out since July 2021) \cite{monfardini2021concerto,catalano_2022}.
Thus, we conclude that the CONCERTO electronics are fully operational and fulfill the technical and scientific requirements.

%%%%%%%%%%%%%%%%%%%%%%%%%%%%%%%%%%%%%%%%%%%%%%%%%%%%%%%%%%%%%%%%%%%%%%%%%%%%%
\section*{Acknowledgments}
This project has received funding from the European Research Council (ERC) under the European’s Horizon 2020 research and innovation programme (grant agreement No 788212), from the Excellence Initiative of Aix-Marseille University-A*Midex, a French "Investissements d’Avenir" programme, and LabEx FOCUS (France).

\bibliography{report_sample_bibtex}
\bibliographystyle{JHEP}

\end{document}